\documentclass{kapproc} 

\RequirePackage{graphicx}%
\RequirePackage{epsf}%
\input{psfig.sty}
\upperandlowercase
\let\footnote\savefootnote
\let\footnotetext\savefootnotetext 
\let\footnoterule\savefootnoterule 
\kluwerbib 

\def\pn{\par\noindent}
\def\sps{\smallskip\pn$\star$}
\def\gtsima{$\; \buildrel > \over \sim \;$}
\def\ltsima{$\; \buildrel < \over \sim \;$}
\def\gsim{\lower.5ex\hbox{\gtsima}}
\def\lsim{\lower.5ex\hbox{\ltsima}}
\def\msun{\hbox{$M_\odot$}}

\def\lsol{L_{\odot}}
\def\lb{\hbox{$L_{\rm B}$}}

\begin{document}


\articletitle{Steeper, Flatter, or Just ``Salpeter''?\
Evidence from Galaxy Evolution and Galaxy Clusters}


\chaptitlerunninghead{Steeper, Flatter, or Just ``Salpeter''?}



 \author{Alvio Renzini}
 \affil{ESO, Garching, Germany}
 \email{arenzini@eso.org}

 \begin{abstract} A single-slope {\it Salpeter} IMF
 overpredicts the stellar $M_*/\lb$ ratio by about a factor of 2,
 which requires the IMF to be flatter below $\sim 1\,\msun$. On the
 other hand a Salpeter IMF for $M\gsim\msun$ predicts an evolution
 with redshift of the fundamental plane of ellipticals in clusters
 which is in agreement with the observations and a formation at
 $z\gsim 3$ for these galaxies. A {\it Salpeter} IMF for $1\lsim M\lsim 40\, 
\msun$ also predicts the observed amount of 
 heavy elements (oxygen and silicon) in clusters of galaxies.

 \end{abstract}

\section{Introduction}
The most direct way of measuring the IMF is by stellar counts, a
technique that can be applied only within a limited distance from
us. On the other hand, much of galaxy properties -- at low as well as
at high redshift -- depend on the IMF, including mass-to-light ratios,
derived star-formation rates, feedback, metal enrichment,
etc. Considering these properties enables us to gather strong
constraints on the IMF for stellar systems and astrophysical
situations that cannot be probed by stellar counts, and provides {\it
sanity checks} for assumptions and theories of the IMF. In this talk I
present the constraints on the IMF slope below $\sim 0.5\msun$,
between $\sim 1$ and $\sim 1.4\,\msun$, and between $\sim 1$ and $\sim
25\,\msun$ that come respectively from the observed values of the
$M_*/\lb$ ratio of local elliptical galaxies, from the redshift
evolution of the fundamental plane, and from the observed mass of
heavy elements in clusters of galaxies. I will also emphasize the role
played by the IMF in our attempts at mapping the cosmic history of
star formation, and the build up of the overall stellar mass content of
the universe. The parameters of the {\it concordance cosmology} are adopted
throughout this paper, i.e., $\Omega_{\rm m}=0.3$; $\Omega_\Lambda
=0.7$, and $H_\circ = 70$.

\section{The Mass-to-Light Ratio of Ellipticals and the IMF Slope Below 
        $\sim 1\,\msun$} 

Fig. 1 shows as a function of age the $M_*/\lb$ ratio for simple stellar
populations (SSP), i.e., assembly of coeaval stars all of the same
(solar) chemical composition.  SSP models are from Maraston (1998),
while the use of other models would give essentially the same
result. The mass-to-light ratio varies wildly depending on the assumed
IMF, which in this case is described by a single slope power law
$\phi(M)\propto M^{-s}$ for $0.1<M<100\,\msun$, 
where $s=2.35$ corresponds to the {\it
Salpeter} IMF (Salpeter 1955). Note that very large mass-to-light
ratios are obtained for either very steep ($s=3.35$) or very flat
($s=1.35$) IMFs. In the former case the total stellar mass is
dominated by the huge number of low mass stars, i.e., the IMF is {\bf
dwarf dominated}. In the latter case instead (at late times) 
 most stellar mass is
locked into remnants: black holes, neutron stars and white
dwarfs, while low mass stars contribute very little to the mass
budget, i.e., the top-heavy IMF leads to a {\bf remnant dominated} mass. Of the
three choices, the {\it Salpeter} slope gives the lowest values of 
$M_*/\lb$. Yet, at an age of $\sim 12$ Gyr, appropriate for
the the bulk stellar population of elliptical
galaxies\footnote{Compelling evidence has accumulated over the years
indicating that the star-formation process in elliptical galaxies was
virtually complete by $z\simeq 3$, corresponding to an age of $\sim
12$ Gyr (for an extensive review see Renzini 1999).}, the $M_*/\lb$
ratio is a factor $\sim 2$ too high compared to the stellar mass of
ellipticals as derived dynamically (e.g. van der Marel 1991). This
excludes any single-slope IMF, hence the IMF must flatten
with respect to the {\it Salpeter} value below $\sim 0.5-0.7\,\msun$, as
indeed indicated by all direct stellar count in Disk and Bulge fields
and in stellar clusters alike, and extensively reported at this
meeting. For example, assuming $s=2.35$ for $M>0.6\,\msun$ and
$s=1.35$ for $M<0.6\,\msun$ the $M_*/\lb$ ratio is reduced by about a
factor of 2, thus bringing SSP models in agreement with the observed
values.

\begin{figure}[ht]
\vskip -2.0 truecm
\centerline{\psfig{file=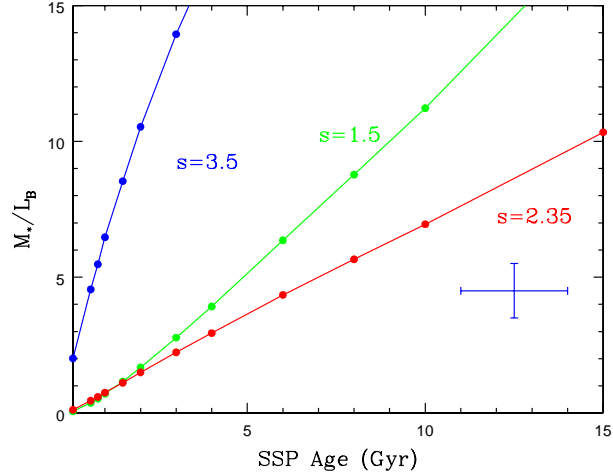,height=3.5in}}
\vskip -5 truemm
\caption{
The time evolution of the stellar $M_*/\lb$ ratio for solar
metallicity simple stellar populations with various slopes ($s$) of the IMF
(synthetic models from Maraston 1998). The cross on the right side of
the figure gives the typical ranges for the $M_*/\lb$ ratio of local
elliptical galaxies and for the ages of their dominant stellar
populations.  }
\vskip -5 truemm
\end{figure}

\section{The Redshift Evolution of the Fundamental Plane, and the IMF Slope 
         for $1\lsim M \lsim 1.4\,\msun$}

As is well known, elliptical galaxies tightly cluster around a plane in
the space having for coordinates the central velocity dispersion
($\sigma$), the effective radius ($R_{\rm e}$) and the effective
surface brightness ($\mu_{\rm e}$) (Djorgovski \& Davis 1987; Dressler
et al. 1987).  This {\it fundamental plane} (FP) combines two
structural/dynamical quantities ($\sigma$ and $R_{\rm e}$) with the
third ($\mu_{\rm e}$) which instead depends on luminosity, hence on
the stellar population content of the galaxy and in particular on its
age.

The slope of the IMF controls the rate of luminosity evolution of a SSP. The
flatter the IMF the faster the luminosity decline past a burst of star
formation.  On the contrary, the steeper the IMF the slower such
decline, as the light from the larger number of low-mass
stars compensates for the progressive death of the more massive
stars. As we look to higher and higher redshift ellipticals, we
therefore expect them to depart from the local FP by an amount that
depends on the slope of the IMF. Thus, if small ellipticals were to have a
different IMF slope w.r.t. large ellipticals, with increasing redshift
one would expect the small ellipticals to depart from the local FP by
a different amount w.r.t. large ellipticals: the FP would {\bf
rotate} with increasing redshift (Renzini \& Ciotti 1993).
Fig. 2 shows that ellipticals in a cluster at $z=0.58$ and another
at $z=0.83$ (corresponding to
a lookback time of $\sim 7$ Gyr, or $\sim 1/2$ the age of the universe)
align parallel to the FP defined by ellipticals in the COMA cluster
(Wuyts et al. 2003): {\bf the FP does not rotate}, hence there is no 
appreciable trend of the IMF with galaxy size, mass, or luminosity.

\begin{figure}[ht]
\vskip -0.6 truecm
\centerline{\psfig{file=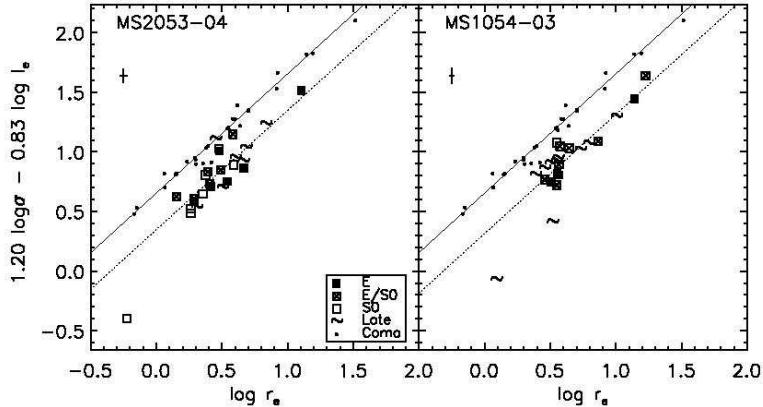,height=2.5in}}
\vskip -5 truemm
\caption{
The fundamental plane of clusters MS2053-04 ($z=0.58$)  and MS1054-03
($z=0.83$) compared to the FP of Coma (Wuyts et al. 2003). Note that the 
slope of the FP appears to remain 
the same with increasing redshift, indicating that the slope of the IMF is
independent of elliptical galaxy size, mass, or luminosity.}
\end{figure}

The shift in the FP shown in Fig. 2 actually corresponds to a change
in the $M/\lb$ ratio of the observed galaxies, and Fig. 3 shows such a
change for the two mentioned clusters, plus a few others at various
redshifts, including one at $z=1.27$ (van Dokkum \& Stanford 2003). 
\footnote{In the $z=1.27$
cluster $\sigma$ was measured for only two
galaxies, and given the errors the slope of the FP could not be reliably 
measured, so only the zero-point shift is shown here.} 
Overplotted is also shown
the evolution with redshift of the $M_*/\lb$ ratio of SSP models 
(from Maraston 1998) having assumed the SSPs formed at $z_{\rm F} =5$. 
The
various $M_*/\lb$ ratios have been normalized to their value at $z=0$, so  
as vto emphasize their relative change with redshift.

\begin{figure}[ht]
\vskip -1.4 truecm
\centerline{\psfig{file=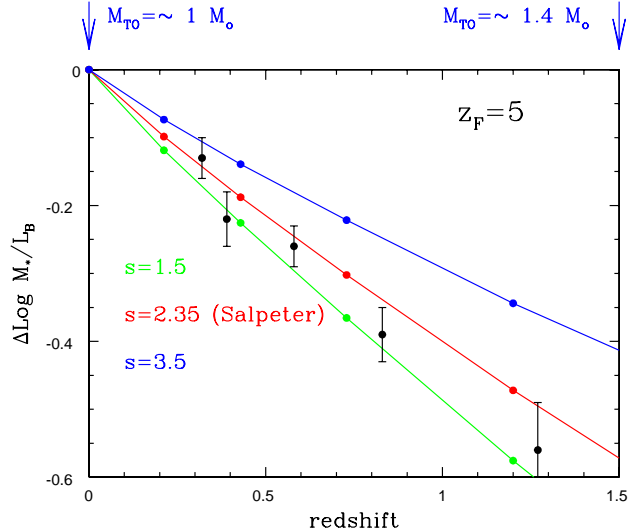,height=3.5in}}
\vskip -5 truemm
\caption{
The evolution with redshift of the $M_*/\lb$ ratio of simple stellar
populations (SSP) of solar metallicity and various IMF slopes,
normalized to its value at $z=0$. A formation redshift $z_{\rm F} =5$
is assumed for the SSPs. The lines refer to $s=1.5$ (top), 2.35 (middle)
and 3.5 (bottom).  The data points refer to the shifts in the
fundamental plane locations for clusters of galaxies at various
redshifts (from van Dokkum \& Stanford 2003). Note that for this
assumed redshift of formation the stellar mass at the main sequence
turnoff is $\sim 1.4\,\msun$ at $z=1.5$ and $\sim \msun$ at $z=0$, as
indicated by the arrows.}
\vskip -3 truemm
\end{figure}

At $z=0$ elliptical galaxies $\sim 12$ Gyr old harbour stellar
populations with $\sim 1\,\msun$ stars at the main sequence turnoff
(MSTO). By $z=1.5$ the precursors of such populations have an age of
only $\sim 3$ Gyr, and correspondingly a higher mass at the MSTO, but
not by much so. Specifically, the MSTO mass at an age of $\sim 3$ Gyr
is $\sim 1.4-1.5\,\msun$, and therefore by following the evolution
of the FP with redshift up to $z\sim 1.5$  (or equivalently of the
mass-to-light ratio) we explore the IMF slope within the rather narrow
mass interval $1\lsim M\lsim 1.4\,\msun$.  In practice, we measure the
slope of the IMF near $M=\msun$.

Note that in Fig. 3 a {\it Salpeter} IMF provides a decent, yet not
perfect fit to the data. On the other hand, at each redshift the
$M_*/\lb$ ratio depends on the assumed age of the SSP, hence on the
assumed redshift of formation $z_{\rm F}$. This means that to some
extent the IMF slope and the formation redshift are
degenerate. Indeed, the {\it Salpeter} IMF provides a better fit to
the FP shifts if one assumes a lower formation redshift, e.g. $z_{\rm
F} =3$, as illustrated in Fig. 4.

Of course, the higher the redshift, the larger the difference between
the $M_*/\lb$ ratios predicted by SSPs with different IMF
slopes. Therefore, measuring the FP shift at the highest possible
redshift would be of great interest. For quite many years the highest
redshift elliptical galaxy known was LBDS 53w091 at $z=1.55$ (Dunlop
et al. 1996; Spinrad et al. 1997), not much beyond the highest
redshift ellipticals in the study by van Dokkum \& Stanford (2003),
which are at $z=1.27$. However, passively evolving elliptical galaxies
have now been identified all the way to $z=1.9$ (Cimatti et al. 2004), for 
which it would be
very interesting to measure the central velocity dispersion, thus
getting the FP parameters, although this may require extremely long 
integrations.

\begin{figure}[ht]
\vskip -2.1 truecm
\centerline{\psfig{file=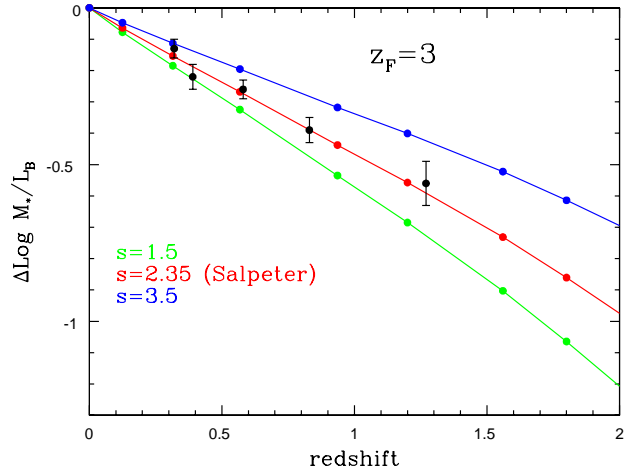,height=3.5in}}
\vskip -5 truemm
\caption{
The same as in Fig. 3 but assuming a formation redshift $z_{\rm
F} =3$.}
\end{figure}

To further illustrate the case of the IMF slope/$z_{\rm F}$
degeneracy, Fig. 5 shows the FP ($M_*/\lb$) shift with redshift for
SSPs with a {\it Salpeter} IMF, solar metallicity, and various
formation redshifts $z_{\rm F}$.  One can conclude that the degeneracy
is strong if the formation redshift of ellipticals were to be
relatively low (e.g., $\sim 2$), but for $z_{\rm F}\gsim 3$ the
degeneracy is quite  mild and the data favor an IMF slope for
$1\lsim M\lsim 1.4\,\msun$ very close to the {\it Salpeter} value.

\section{The Metal Content of Galaxy Clusters, and the IMF Slope for
 $1\lsim M\lsim 25\,\msun$}

In clusters of galaxies most of the light comes from early-type
galaxies dominated by old stellar populations, i.e., from
$\sim 1\,\msun $ stars. Meanwhile, one can also measure the amount
(mass) of metals contained in the intracluster medium and in the
stars. As most metals are produced by Type II supernovae, the total
metal mass in a cluster of galaxies is proportional to the number
of massive stars ($M\gsim 8\,\msun$) that have exploded in the far
past, therefore dispersing the metals we see today\footnote{Iron is
the element whose abundance is most reliably measured both in the ICM
and in galaxies. However, its production may be dominated by Type
Ia supernovae, whose progenitors are binary stars. Given the 
complexities introduced by the binary nature of the iron producers, 
this element is less useful for the
determination of the IMF slope.}.  It follows that the {\it
metal-mass-to-light ratio} of galaxy clusters (e.g., Renzini 2003) provides
a measure of the number ratio of massive to $\sim \msun$ stars, i.e.,
of the IMF slope between $\sim 1$ and $\sim 25\,\msun$ (or
more)\footnote{With a Salpeter IMF the {\it typical} stellar
mass for nucleosynthesis yields is $\sim 25\,\msun$, i.e., the overall
yield can be approximated by the total number of Type II supernovae
times the nucleosynthesis products of a 25$\,\msun$ star.} for the
stellar populations that -- when young -- have produced the metals and now
-- that are old -- produce the light we see today.

\begin{figure}[ht]
\vskip -1.4 truecm
\centerline{\psfig{file=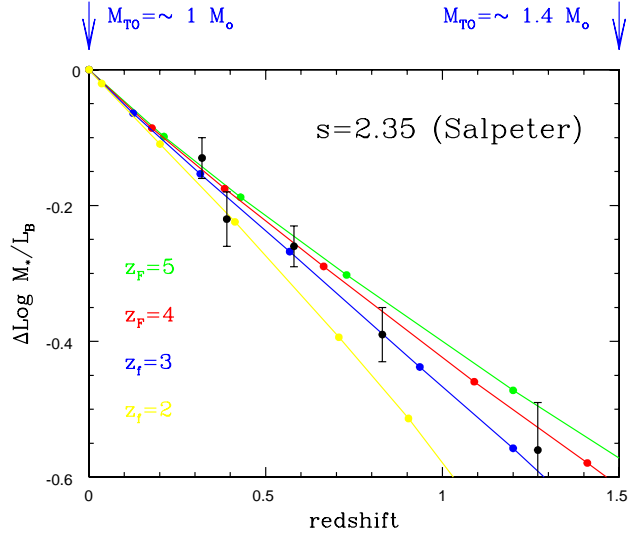,height=3.5in}}
\vskip -5 truemm
\caption{The same as in Fig. 3 for the Salpeter IMF
and various values of the formation redshift $z_{\rm
F}$, from $z_{\rm F}=5$ (top) to $z_{\rm F} =2$ (bottom).}
\end{figure}

The IMF of a passively evolving SSP can be written as:

\begin{equation}
\phi(M)=a(t,Z)\lb M^{-s},
\end{equation}
where $\lb$ is its $B$-band luminosity and $a(t,Z)$ is a (slow)
function of the SSP age and metallicity. For example, for $Z=Z_\odot
=0.02$ one has $a(t) = 1.67$ and 2.51, respectively for $t=10$ and 15 Gyr
(Maraston 1998) (with $M$ and $\lb$ 
in solar units). More recent models
(Maraston 2004, in preparation) give $a(t)=1.68$ and 2.8 for the same $(t,Z)$
and $a(t)=2.50$ and 3.94 for $Z=2Z_\odot$. For $t=12$ Gyr (the minimum age
for $z_{\rm F}\gsim 3$) the new Maraston models give $a(Z)=2.22$ and 3.12,
respectively for $Z=Z_\odot$ and $Z=2Z_\odot$.

Therefore, the metal-mass-to-light
ratio for the ``X'' element can be calculated in a straightforward manner from:
\begin{equation}
{M_{\rm X}\over\lb} = a(t,Z)\int_8^{40}m_{\rm X}(M)M^{-s}dM,
\end{equation}
\begin{figure}[ht]
\vskip -1.8 truecm
\centerline{\psfig{file=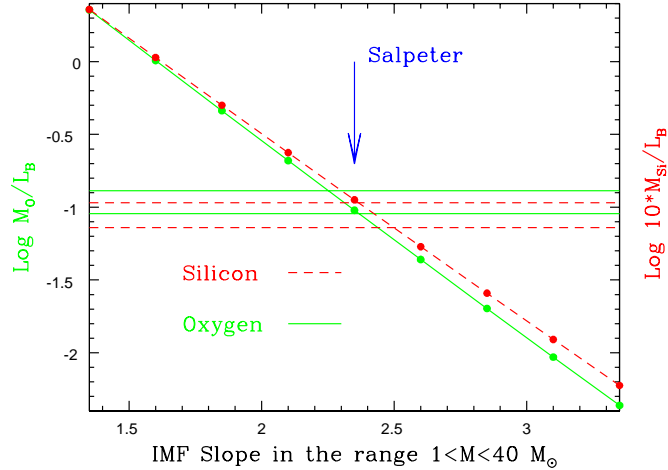,height=3.5in}}
\vskip -8 truemm
\caption{
The Oxygen- and the Silicon-Mass-to-Light Ratios as a function of the 
IMF slope as calculated from Eq. (2) with$a(t,Z)=3$ and
  using the nucleosynthesis 
prescription of Woosley \& Weaver (1995). The horizontal lines show the 
observed average values of this ratios in clusters of galaxies, and their 
uncertainty range.}
\end{figure}
\noindent
where $m_{\rm X}(M)$ is the mass of the element ``X'' which is produced 
by a star of mass $M$. Adopting $a(t,Z)=3$ and $m_{\rm X}(M)$
 for oxygen and silicon from
Woosley and Weaver (1995) and integrating Eq. (2) one obtains the 
{\it oxygen-} and the {\it silicon-mass-to-light ratios},
which are shown in Fig. 6 as a function of the IMF slope.
As expected, the $M_{\rm O}/\lb$ and  $M_{\rm Si}/\lb$ ratios are extremely 
sensitve to the IMF slope. The values observed in clusters of galaxies
(ICM plus galaxies) are $\sim 0.1$ and $\sim 0.01\,\msun/\lsol$, respectively 
for oxygen and silicon\footnote{These values result from averaging over the
reported values for individual clusters with different ICM temperature, and 
take into 
account that $\sim 10-30\%$ of the stellar mass in clusters is not bound to 
individual galaxies (Arnaboldi et al. 2003; Gal-Yam et al. 2003).} (e.g., 
Finoguenov et al. 2003; Portinari et al. 2004, and references therein). 
These empirical values are reported in Fig. 6, which at the meeting was 
presented to Ed Salpeter as a Happy-Birthday postcard. Indeed, with the {\it
Salpeter} IMF slope ($s=2.35$) the standard explosive nucleosynthesis
from Type II supernovae produces just the right amount of oxygen and silicon
to account for the observed $M_{\rm O}/\lb$ and  $M_{\rm Si}/\lb$ ratios
in cluster of galaxies, having assumed that most of the cluster $B$-band light
comes from $\gsim 12$ Gyr stellar populations.

Fig. 6 also shows that with $s=1.35$ such a {\it top heavy} IMF (in
various circumstances invoked to ease discrepancies between theories and
observations) would overproduce metals by more than a factor
of 20. This is indeed the change one expects in $M_{\rm O}/\lb$,
$M_{\rm Si}/\lb$, etc. for a $\Delta s=1$ when considering that the
light $\lb$ is provided by $\sim\msun$ stars and the metals by $\sim
25\,\msun$ stars.

\section{Masses and Star Formation Rates of High Redshift Galaxies}

As is well known, all SFR indicators (UV continuum, H$\alpha$,
sub-mm, etc.) measure the formation rate of massive stars. To
derive the total SFR an IMF (slope and shape) has to be
assumed. Similarly, an IMF must be assumed to derive the total stellar mass
of a galaxy from its spectral energy distribution (SED) and
luminosity. These are indeed determined by the number of stars producing
the bulk of the light (which are typically in a quite narrow range of
masses), while remnants and dwarfs produce little light but may
contribute a major fraction of the mass.  It follows that the time
integral of the measured cosmic
SFR($z$) (in $\msun$yr$^{-1}$Mpc$^{-1}$)
should agree with the measured stellar mass
density $\rho_*(z)$ (in $\msun$Mpc$^{-1}$), and in turn at $z=0$ this should
agree with its dynamical estimate. Clearly, such a general
agreement needs to be achieved with the {\it right} value of the IMF.
Fig. 7 shows a recent attempt in this direction (adopting a Salpeter
IMF, Fontana et al. 2004) but error bars are still too large to allow
a definitive answer. However, several galaxy surveys (with HST, SST,
VLT, etc. etc.) are now well
under way, and extremely rapid progress is expected in this field.

\begin{figure}[ht]
\centerline{\psfig{file=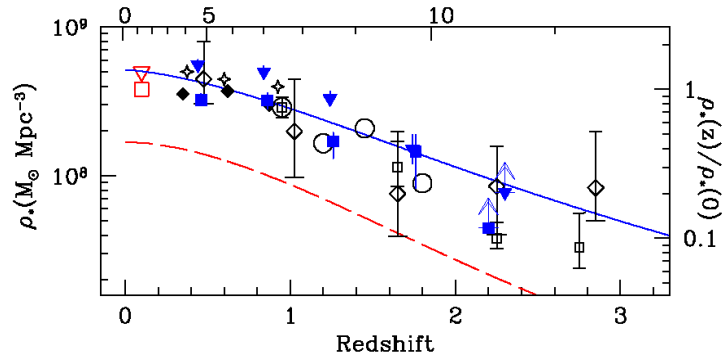,angle=-90,height=2in}}
\caption{The cosmic evolution of the comoving density of the overall
 mass in stars from various current surveys, compared with the time integral
of the cosmic star-formation rate density, uncorrected (dashed line) and 
corrected for dust attenuation effects (solid line) (from Fontana et al. 2004). 
}
\end{figure}

\section{Summary}
\sps
The IMF of elliptical galaxies is close to {\it Salpeter} except below
\par $\sim 0.6\,\msun$, where it flattens.
\sps
The fundamental plane (FP) of ellipticals does not  rotate
with redshift, \par implying that the IMF is independent of galaxy
luminosity (mass).
\sps
The FP shift with redshift is consistent with a {\it Salpeter} IMF
near \par $M\sim \msun$ \par and a formation at $z\gsim 3$ for the
bulk of stars in ellipticals.
\sps
The overall metal content of galaxy clusters (ICM+galaxies) is well \par
reproduced by a {\it Salpeter} IMF and classical nucleosynthesis.
\sps
Very rapid progress is expected in determining the cosmic history of
\par
star formation along with the cosmic build-up of stellar mass and a \par
consistent IMF.
\sps
The IMF may not be universal (who knows??) but the one of which \par
this meeting celebrates 
the 50th anniversary doesn't quite show its \par age (!).
%
\medskip\pn
{\it Acknowledgments.} I would like to thank Claudia Maraston for having 
provided her most recent models in advance of publication, and the staff of
the Carnegie Observatories (Pasadena) for their kind hospitality during the 
time when this paper was written and set up.

\end{document}